\documentclass[a4paper,fleqn,usenatbib]{mnras}

\usepackage{newtxmath}

\usepackage{graphicx}	
\usepackage{amsmath}	
\usepackage{amssymb}	
\usepackage{amsfonts}
\usepackage{multirow}
\usepackage{threeparttablex}
\usepackage{xcolor}
\colorlet{RED}{red}    

\usepackage[color=white]{todonotes}

\title[X-ray properties of $z>6.5$ quasars]{X-ray properties of $\mathbf{z\ga6.5}$ quasars}

\author[E. Pons et al.]
{\parbox{\textwidth}
{E. Pons$^{1,2}$\thanks{E-mail: epons@ast.cam.ac.uk},
R. G. McMahon$^{1,2}$,
M. Banerji$^{1,2}$,
and S. L. Reed$^{3}$
}
\\ \\
$^{1}$Institute of Astronomy, University of Cambridge, Madingley Road, Cambridge CB3 0HA, UK\\
$^{2}$Kavli Institute for Cosmology, University of Cambridge, Madingley Road, Cambridge CB3 0HA, UK\\
$^{3}$Department of Astrophysical Sciences, Princeton University, 4 Ivy Lane, Princeton, NJ 08544, US
}

\date{Accepted XXX. Received YYY; in original form ZZZ}

\pubyear{2019}

\begin{document}
\label{firstpage}
\pagerange{\pageref{firstpage}--\pageref{lastpage}}
\maketitle


\begin{abstract}
We present \emph{XMM-Newton} X-ray observations and analysis of three DES $z>6.5$ quasars (VDES J0020$-$3653 at $z=6.824$, VDES J0244$-$5008 at $z=6.724$ and VDES J0224$-$4711 at $z=6.526$) and six other  quasars with $6.438 < z < 6.747$ from the \emph{XMM-Newton} public archive. Two of the nine quasars are detected at a high ($>$4$\sigma$) significance level: VDES J0224$-$4711(z=6.53) at $9\sigma$ and PSO J159$-$02 ($z=6.38$) at $8\sigma$. They have a photon index of $\Gamma=1.82^{+0.29}_{-0.27}$ and $\Gamma=1.94^{+0.31}_{-0.29}$ respectively, which is consistent with the mean value of $\sim1.9$ found for quasars at all redshifts. The rest-frame $2-10$ keV luminosity of VDES J0224$-$4711, is $L_{2-10\mathrm{keV}} = (2.92\pm0.43)\times10^{45}\;\mathrm{erg\;s^{-1}}$, which makes this quasar one of the most X-ray luminous quasars at $z>5.5$ and the most X-ray luminous quasar at $z>6.5$, with a luminosity 6 times and 2.5 times larger than ULAS J1120+0641 ($z=7.08$) and ULAS J1342+0928 ($z=7.54$) respectively. 
The X-ray-to-optical power-law slopes of the nine quasars are consistent with the previously observed anti-correlation of $\alpha_{ox}$ with UV luminosity  $L_{2500\mathrm{\AA}}$. We find no evidence for evolution of $\alpha_{ox}$ with redshift when the anti-correlation with UV luminosity is taken into account.
Similar to previous studies at z$\sim$6 we have found remarkably consistent X-ray spectral properties between low (z$\sim$1) and high-redshift quasars. Our results add further evidence to the picture that the observable properties of high luminosity quasars over the UV to X-ray spectral region have not evolved significantly from $z\sim7$ to the present day and that quasars comparable to local versions existed 800 million years after the big bang.
\end{abstract}




\section{Introduction}

\begin{table*}
   \caption{XMM observations log for the DES and new $z>6.4$ X-ray quasars}
   \label{tab:xmm_log}
   \begin{threeparttable}
      \begin{tabular}{lrrccccccc}
\hline
  \multicolumn{1}{c}{\multirow{2}{*}{Object}} &
  \multicolumn{1}{c}{\multirow{2}{*}{R.A.}} &
  \multicolumn{1}{c}{\multirow{2}{*}{Dec}} &
  \multicolumn{1}{c}{\multirow{2}{*}{$z$}} &
  \multicolumn{1}{c}{$N_{H,Gal}$} &
  \multicolumn{1}{c}{\multirow{2}{*}{Obs. Date}} &
  \multicolumn{1}{c}{$t_{exp}$\tnote{a}} &
  \multicolumn{1}{c}{XMM} &
  \multicolumn{1}{c}{\multirow{2}{*}{PI}} &
  \multicolumn{1}{c}{\multirow{2}{*}{Ref\tnote{b}}}\\
   & & & & \multicolumn{1}{c}{($10^{20}\;\rm{cm^{-2}}$)} & & \multicolumn{1}{c}{(ks)} &\multicolumn{1}{c}{OBSID} & &\\
\hline
  VDES J0020$-$3653 & 5.13113 & -36.89494 & 6.834 & 1.53 & 2018 May 14 & 24.8 & 0824400101 & Pons E.& 2/2\\
  VDES J0244$-$5008 & 41.00425 & -50.14825 & 6.724 & 2.55 & 2018 May 16 & 17.0 & 0824400201 & Pons E. & 2/2\\
  VDES J0224$-$4711 & 36.11058 & -47.19149 & 6.526 & 1.66 & 2018 May 25 & 26.1 & 0824400301 & Pons E. & 1/2\\
  \hline \hline
  PSO J338.2298+29.5089 & 338.22975 & 29.50897 & 6.658 & 6.17 & 2017 Nov. 26 & 23.7 & 0803160301 & Schartel N. & 4/4\\
  VIK J0109$-$3047 & 17.47138 & -30.79064 & 6.747 & 2.32 & 2017 Dec. 06  & 6.0 & 0803160201 & Schartel N. & 3/8\\
  PSO J159.2257$-$02.5438 & 159.22579 & -2.54386 & 6.38 & 4.74 & 2017 Dec. 22 & 19.9 & 0803161101 & Schartel N. & 5/5\\
  VIK J0305$-$3150 & 46.32050 & -31.84886 & 6.605 & 1.42 & 2018 Jan. 05 & 22.0 & 0803160401 & Schartel N. & 3/8\\
  PSO J036.5078+03.0498 & 36.50779 & 3.04983 & 6.541 & 3.04 & 2018 Jan. 16 & 16.8 & 0803160501 & Schartel N. & 4/4\\
  CFHQS J0210$-$0456 & 32.55496 & -4.93914 & 6.438 & 1.88 & 2018 Feb. 02 & 8.4 & 0803160701 & Schartel N. & 6/7\\
  \hline
  \end{tabular}

      \begin{tablenotes}
         \small
         \item [a] EPIC net exposure times after high particle background filtering. Mean between the pn and MOS detectors.
         \item [b] Discovery / redshift references: (1) \citet{Reed17}, (2) \citet{Reed19}, (3) \citet{Venemans13}, (4) \citet{Venemans15}, (5) \citet{Banados16}, (6) \citet{Willott10a}, (7) \citet{Willott10b} and (8) \citet{DeRosa14}.
    \end{tablenotes}
   \end{threeparttable}
\end{table*}

High-redshift quasars with $z\ga6.5$ are important probes of the Universe during the epoch of Reionization providing information about the formation, growth and evolution of supermassive black holes (SMBH) and their host galaxies: indeed the presence of $10^9\;\mathrm{M_{\odot}}$ at $z>6.5$, 800 million years after the big bang is a challenge for models of black-hole 
formation and growth \citep{Volonteri10, Alexander14, Trakhtenbrot17}. In the last decade, about 30 $z\ga 6.5$ quasars have been discovered mainly from  large-area optical and near-infrared imaging survey such as the CFHT Legacy Survey \citep[CFHTLS;][]{Willott10a}, Pan-STARRS \citep{Venemans15, Tang17, Mazzucchelli17}, the Dark Energy Survey \citep[DES;][]{Reed17, Reed19, Yang19}, the Dark Energy Spectroscopic Instrument Legacy Imaging Surveys \citep[DELS;][]{Wang17, Wang18, Wang19}, the Subaru HSC-SSP survey \citep{Matsuoka16, Matsuoka18a, Matsuoka18b, Matsuoka19}, VISTA \citep{Venemans13, Pons19} and the UKIRT InfraRed Deep Sky Surveys-Large Area Survey \citep[ULAS;][]{Mortlock11, Banados18}.


X-ray observations are a very powerful tool to understand quasars providing information on the region close to the central SMBH. However only two $z > 6.5$ quasars have published  observation at X-rays; ULAS J1120+0641 at $z=7.1$ \citet{Page14} and ULAS J1342+0928 at $z=7.5$ \citet{Banados18b}. 
X-ray studies of high-redshift ($z\sim6$) quasars have shown that their optical and X-ray properties are similar to those of low redshift quasars with similar average X-ray spectral photon index \citep[see for example][]{Farrah04, Page05, Nanni17} and comparable luminosity at $\lambda=2500\mathrm{\AA}$. Also, no dependence of the optical-to-X-ray slope ($\alpha_{ox}$) with redshift have been found up to $z\sim6$ \citep{Brandt02, Steffen06} - the apparent correlation being only an artifact from the dependence of $L_{2500\mathrm{\AA}}$ with the redshift \citep{Steffen06, Lusso10} - suggesting that the central X-ray energy source of quasars does not evolve strongly over cosmic time. Furthermore, previous work \citep{Steffen06, Lusso10} has found evidence that $\alpha_{ox}$ anti-correlates with the quasar UV luminosity $L_{2500\mathrm{\AA}}$, 
with more luminous quasars having more negative values of $\alpha_{ox}$ (i.e. $\alpha_{ox}$ i.e. relativle less X-ray emission as $L_{2500\mathrm{\AA}}$ increases). 
This indicates that for powerful quasars the X-ray emission produced by the hot corona is weaker relative to the UV and optical emission from the disk.

In order to improve our knowledge of the X-ray properties of $z > 6.5$ quasars, in this paper we report \emph{XMM-Newton} observations and analysis of a sample of nine quasars with $z \ga 6.5$;
three quasars from the Dark Energy Survey (DES) at $z>6.5$, VDES J0020$-$3653 ($z=6.824$), VDES J0244$-$5008  ($z=6.724$) VDES J0224$-$4711 ($z=6.526$) and of six other  quasars with $6.438 < z < 6.747$ from the \emph{XMM-Newton} public archive in Section \ref{sec:data}.
The data analysis of these sources is presented in Section \ref{sec:results} and and finally in Section \ref{sec:disc}, we discuss the X-ray and optical properties of the nine quasars 
in our sample compared to previous X-ray observations of high-redshift quasars.
We assume a flat cosmology with $H_0=70\;\mathrm{km\;s^{-1}}$, $\Omega_M=0.3$ and $\Omega_{\Lambda}=0.7$.

\section{XMM observations and data reduction}
\label{sec:data}

In this paper we present the \emph{XMM-Newton}  of three spectroscopically confirmed high-redshift ($z>6.5$) quasars recently 
discovered using observations from the Dark Energy Survey (DES) \citet{Reed17, Reed19} though colour 
selection and spectral energy distribution fitting of DES optical, VISTA near-infrared and WISE mid-infrared photometry. \emph{XMM-Newton} observing time were allocated through the call of proposal for period AO-17, and they were observed in 2018 May with 28.3ks, 28.0ks and 35.9ks exposure for VDES J0020$-$3653, VDES J0244$-$5008 and VDES J0224$-$4711 respectively. In addition we present the analysis of six other quasars with $6.44<z<6.75$ with X-ray observations in the \emph{XMM-Newton} public archive. These six quasars were also selected from wide field optical and near-IR surveys i.e. CFHTLS \citep{Willott10a}, Pan-STARRS1 \citep{Banados16, Venemans15} and VISTA VIKING \citep{Venemans13}.  They have been observed between 2017 November and 2018 February for a total observing time of 23 to 33ks. The redshifts as well as the X-ray observations log for the full sample of 9 z$\ga$6.4 quasars are reported in Table \ref{tab:xmm_log} and J-band continuum magnitudes are given in Table \ref{tab:XrayOpt}.



The X-ray observational data were obtained with the European Photon Imaging Camera \citep[EPIC;][]{Struder01, Turner01}, which consists of one pn and two MOS cameras, and was operated in full-frame mode with thin filters for all the observations. The data were processed using the standard \emph{XMM-Newton Science Analysis Software} (\texttt{SAS}) with version 17.0.0. We excluded time intervals affected by high particle background through inspection of the light curves in the 10$-$12 keV energy range, resulting of net exposure times between 17.0 and 26ks (see Table \ref{tab:xmm_log}). To construct events files we used the standard EPIC events pattern selection; i.e. 0$-$12 and 0$-$4 for MOS and pn respectively. We then produced images in five energy bands: 0.2$-$10, 0.2$-$0.5, 0.5$-$2.0, 2.0$-$5.0 and 5.0$-$10.0 keV. 

To check if the sources were detected and measure the X-ray flux, we performed a simultaneous EPIC (pn+MOS) sources detection in each band using the \texttt{SAS} task \texttt{edetect\_chain} with a detection likelihood threshold of 10 (corresponding to $4\sigma$). We also used the \texttt{SAS} task \texttt{eregionanalyse} to compute fluxes or perform $3\sigma$ upper limit estimations in each energy bands on the pn and MOS images, by extracting the number of counts in a circular region of 10$\arcsec$ radius. To get an accurate estimation of the number of counts, we used the exposure map created by the \texttt{edetect\_chain} task and we selected a background circular region adjacent to the target and on the same chip, free of sources, with a larger radius.

In order to convert the count rates to flux for our different energy ranges and detectors, we computed energy conversion factors (ecf) using WebPIMMS, converting from a XMM-MOS/pn thin count rate to a flux assuming a power-law spectral shape with a slope of 1.9 modified by galactic absorption. The EPIC parameters (i.e. source counts, count rates and fluxes, HR) correspond to the weighted mean of the values obtained for the pn, MOS1 and MOS2 cameras. When the resulting flux values are smaller than their 3$\sigma$ errors the $3\sigma$ upper limits are given.

In the case where an X-ray source is detected at greater than $4\sigma$ (i.e. VDES J0224$-$4711 and PSO J159$-$02, see Section \ref{sec:results}), we extracted a spectrum from a 
15$\arcsec$/11$\arcsec$ radius region around the target position
in the pn/MOS detectors. The background spectrum was extracted on the same region that the one used by the \texttt{eregionanalyse} task. We then combined the pn and MOS spectra into a single EPIC spectrum using the \texttt{epicspeccombine}, with corresponding background spectra and response matrix also combined. For the resulting EPIC spectra, we grouped the data with only one count per bin, due to low number of counts.

\section{Results}
\label{sec:results}

\begin{figure}
    \includegraphics[width=\columnwidth]{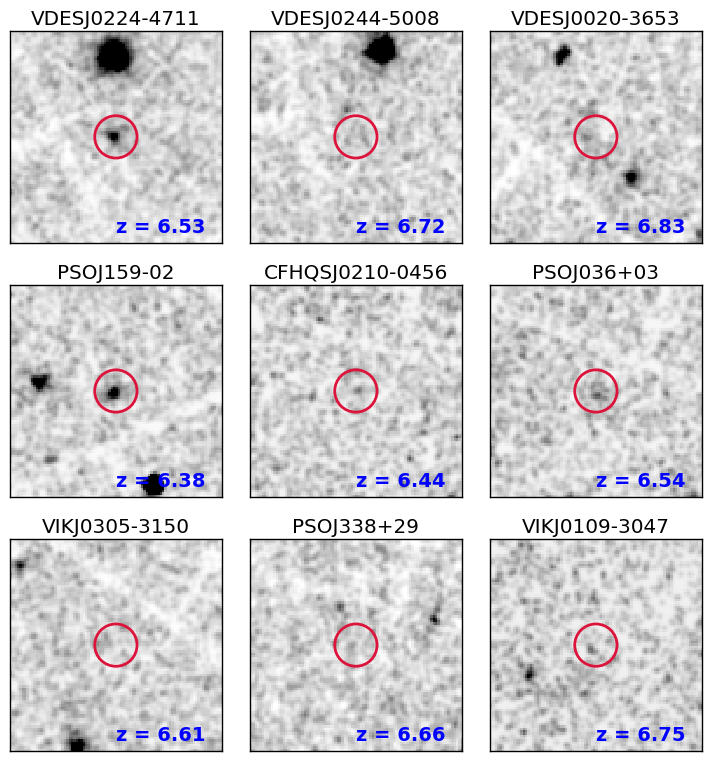}
    \caption{Full band (0.2$-$10 keV) EPIC 100$\arcsec$ mosaics for the three DES $z>6.5$ quasars (top row) and 6 newly analysed $z>6.4$ quasars. Red circles are 10$\arcsec$ radius, which correspond to the extraction radius. The images has been smoothed with a Gaussian function of $\sigma=$ 1.0 pixel.} 
    \label{fig:cutouts}
\end{figure}

From our three DES targets, only one has been clearly detected in the X-rays: VDES J0224$-$4711. Its detected net counts in the full 0.2$-$10 keV energy range is 174, corresponding to a flux of $\sim 12\times10^{-15}\;\mathrm{erg\;cm^{-2}\;s^{-1}}$ with a detection significance of $9\sigma$; the X-ray source is clearly visible on the image Figure \ref{fig:cutouts}. It is also detected individually in the 0.5$-$2.0 keV energy range with 118 net counts and a detection significance of $8\sigma$ (see Table \ref{tab:xmm-vdes_xray}). The separation between the optical position of this quasar and the \emph{XMM} detection is only 2.52$\arcsec$, with the uncertainty on the X-ray position beeing less than 1$\arcsec$ ($\sim 0.81\arcsec$).

The two other DES quasars, VDES J0244$-$5008 and VDES J0020$-$3653, are not detected using the \texttt{edetect\_chain} task. From the \texttt{eregionanalyse} task, VDES J0020$-$3653 have about 100 net counts in the 0.2$-$10keV energy range, corresponding to a flux of $\sim 6\times10^{-15}\;\mathrm{erg\;cm^{-2}\;s^{-1}}$. Finally, VDES J0244$-$5008 have only 45 counts in the full energy range and a flux of $\sim 5\times10^{-15}\;\mathrm{erg\;cm^{-2}\;s^{-1}}$.

\begin{figure}
    \includegraphics[width=\columnwidth, trim={0.5cm 1.5cm 3cm 2.5cm}, clip]{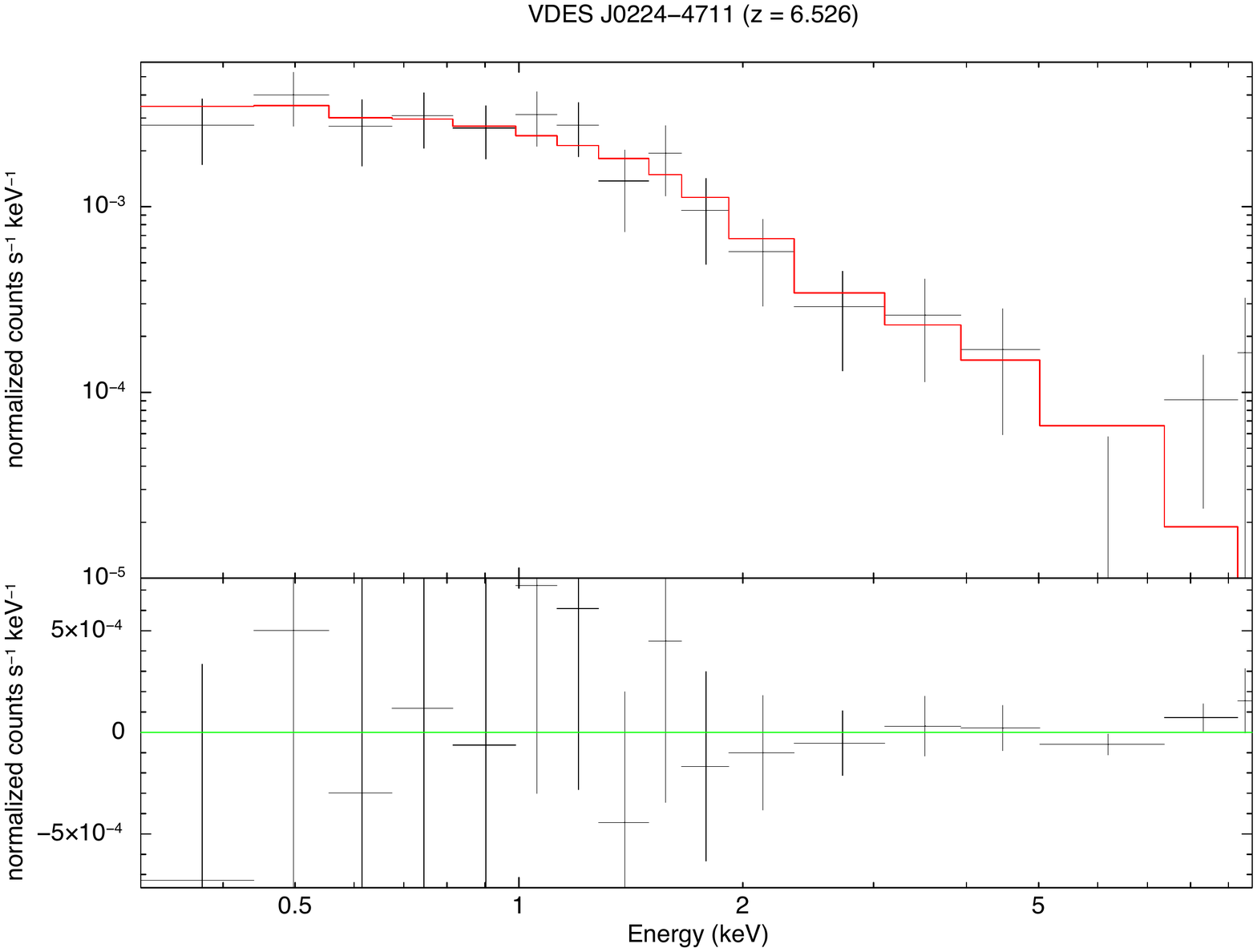}
    \caption{\emph{Upper panel}: \emph{XMM-Newton} spectrum of VDES J0224$-$4711. The fit with a power law component modified by Galactic absorption is shown by the red line. \emph{Lower panel}: Residuals between the data and the best fit model. For a better visualisation adjacent bins have been combined until they have a significant detection of at least $20\sigma$ (\texttt{XSPEC rebin} command). This is for plotting purpose only and has no influence on the fit.}
    \label{fig:spec}
\end{figure}

The EPIC spectrum for VDES J0224$-$4711 is shown in Figure \ref{fig:spec}. We fit it using \texttt{XSPEC}  with a single power-law model (\texttt{zpo}) modified by Galactic absorption (\texttt{tbabs}). Due to the low number of counts, we used the Cash statistic \cite{Cash79}. With the Galaxy absorption component kept fix, we found a photon index on $\Gamma = 1.82^{+0.29}_{-0.27}$ with $\mathrm{C-stat} = 135.5$ for 161 degrees of freedom and get a flux in the 2$-$10keV energy range of $f_{2-10kev}=(7.38\pm1.09)\times10^{-15}\;\mathrm{erg\;cm^{-2}\;s^{-1}}$. Such a value of $\Gamma$ is consistent with what is observed for quasars in the low \citep{Page05, Piconcelli05, Just07} and high \citep{Shemmer05, Nanni17} redshift Universe.

For the three DES quasars, the EPIC net source counts (i.e. after background subtraction) and fluxes in the full (0.2$-$10 keV) energy range in in the 0.2$-$0.5, 0.5$-$2.0, 2.0$-$5.0 and 5.0$-$10.0 keV energy ranges are given in Table \ref{tab:xmm-vdes_xray}. The full band \emph{XMM} 100$\arcsec$ EPIC mosaic of the DES $z>6.5$ quasars are shown in Figure \ref{fig:cutouts} (top row).

\begin{table}
  {\centering
  \caption{X-ray net counts and fluxes for the DES quasars}
  \label{tab:xmm-vdes_xray}
  \begin{tabular}{l|rrr}
\hline
  \multicolumn{1}{c|}{\multirow{2}{*}{Energy range}} &
  \multicolumn{1}{c}{\multirow{2}{*}{Net counts}} &
  \multicolumn{1}{c}{Flux} &
  \multicolumn{1}{c}{\multirow{2}{*}{Det.}} \\
   & & \multicolumn{1}{c}{($10^{-15}\;\mathrm{erg\;cm^{-2}\;s^{-1}}$)} &  \\
\hline
  \multicolumn{4}{c}{VDES J0224$-$4711 ($z=6.526$)}\\
  \hline
  \bf{0.2 - 10 keV} & $173.3\pm 20.3$ & $11.96\pm1.39$ & $9\sigma$\\
  0.2 - 0.5 keV & $<23.8$ & $<1.76$ & \\
  0.5 - 2.0 keV & $118.1\pm15.4$ & $5.08\pm0.65$ & $8\sigma$\\
  2.0 - 5.0 keV & $29.2\pm8.8$ & $3.83\pm1.16$ & \\
  5.0 - 10 keV & $<16.5$ & $<22.48$ & \\
  \hline
  \multicolumn{4}{c}{VDES J0244$-$5008 ($z=6.724$)}\\
  \hline
  \bf{0.2 - 10 keV} & $<34.6$ & $<8.8$ & \\
  0.2 - 0.5 keV & $<13.6$ & $<2.71$ & \\
  0.5 - 2.0 keV & $<24.0$ & $<4.50$ & \\
  2.0 - 5.0 keV & $<7.1$ & $<6.49$ & \\
  5.0 - 10 keV & $<4.1$ & $<31.41$ & \\
  \hline
  \multicolumn{4}{c}{VDES J0020$-$3653 ($z=6.834$)}\\
  \hline
  \bf{0.2 - 10 keV} & $36.0\pm11.7$ & $2.68\pm0.87$ & \\
  0.2 - 0.5 keV & $<18.2$ & $<2.13$ & \\
  0.5 - 2.0 keV & $<26.0$ & $<2.39$ & \\
  2.0 - 5.0 keV & $<5.6$ & $<5.47$ & \\
  5.0 - 10 keV & $<14.1$ & $<24.86$ & \\
\hline
\end{tabular}
}
  
  \textbf{Note}: The upper limits quoted in this table correspond to the $3\sigma$ upper limits and are used only if the nets counts or fluxes are smaller than their $\sigma$ error. But only the quasars with a given value in the "Det" column have an X-ray detection.
\end{table}


Among the six other quasars, only PSO J159.2257$-$02.5438 ($z = 6.38$) has been detected with 127 counts in the 0.2$-$10 keV band, corresponding to a flux of $\sim 11\times10^{-15}\;\mathrm{erg\;cm^{-2}\;s^{-1}}$, and a detection significance of $8\sigma$. It is also detected individually in the 0.5$-$2.0 keV and 2.0$-$5.0 keV energy bands with 84 counts ($7\sigma$ significance) and 27 counts ($4\sigma$ significance) respectively. As for VDES J0224-4711, the separation between the optical and X-ray positions is quite small ($\sim 1.30\arcsec$) and the error on the \emph{XMM} position is only $0.83\arcsec$. From fitting its spectrum with a single power-law modified by Galactic absorption, we get a photon index of $\Gamma=1.94^{+0.31}_{-0.29}$ ($\mathrm{C-stat} = 117.2$ for 115 degrees of freedom), similar to what we found for VDES J0224$-$4711, and we obtained a 2$-$10 keV flux of $f_{2-10kev}=(5.52\pm1.72)\times10^{-15}\;\mathrm{erg\;cm^{-2}\;s^{-1}}$. For the undetected 5 other quasars we provide estimation of the net source counts and fluxes from the \texttt{eregionanalyse} task. They have less than 35 counts in the full band, for three of them we even only get upper limits. The EPIC net source counts and fluxes in the full (0.2$-$10 keV) energy range in in the 0.2$-$0.5, 0.5$-$2.0, 2.0$-$5.0 and 5.0$-$10.0 keV energy ranges obtained for these quasars are summarised in Table \ref{tab:xmm-schartel_xray}. The full band \emph{XMM} 100$\arcsec$ EPIC mosaic of these 6 quasars are shown in Figure \ref{fig:cutouts} (rows 2 and 3).

For the two X-ray detected quasars in this work, the source counts and fluxes presented in Tables \ref{tab:xmm-vdes_xray} and \ref{tab:xmm-schartel_xray} are from the \texttt{edetect\_chain} task.

\begin{table}
  {\centering
  \caption{X-ray net counts and fluxes for the new $z>6.4$ X-ray quasars}
  \label{tab:xmm-schartel_xray}
  \begin{tabular}{l|rrr}
\hline
  \multicolumn{1}{c|}{\multirow{2}{*}{Energy range}} &
  \multicolumn{1}{c}{\multirow{2}{*}{Net counts}} &
  \multicolumn{1}{c}{Flux} &
  \multicolumn{1}{c}{\multirow{2}{*}{Det.}} \\
   & & \multicolumn{1}{c}{($10^{-15}\;\mathrm{erg\;cm^{-2}\;s^{-1}}$)} &  \\
\hline
  \multicolumn{4}{c}{PSO J159.2257$-$02.5438 ($z=6.38$)}\\
  \hline
  \bf{0.2 - 10 keV} & $127.1\pm15.8$ & $10.82\pm1.35$ & $8\sigma$ \\
  0.2 - 0.5 keV & $<24.4$ & $<4.4$ & \\
  0.5 - 2.0 keV & $83.8\pm11.8$ & $4.11\pm0.58$ & $7\sigma$\\
  2.0 - 5.0 keV & $26.5\pm7.2$ & $3.90\pm1.13$ & $4\sigma$\\
  5.0 - 10 keV & $<8.4$ & $<26.72$ & \\
 \hline
 \multicolumn{4}{c}{CFHQS J0210$-$0456 ($z=6.438$)}\\
  \hline
  \bf{0.2 - 10 keV} & $<16.2$ & $<14.39$ &  \\
  0.2 - 0.5 keV & $<6.7$ & $<5.77$ & \\
  0.5 - 2.0 keV & $<6.1$ & $<5.53$ & \\
  2.0 - 5.0 keV & $<2.6$ & $<11.81$ & \\
  5.0 - 10 keV & $<11.2$ & $<77.09$ & \\
 \hline

  \multicolumn{4}{c}{PSO J036.5078+03.0498 ($z=6.541$)}\\
  \hline  
  \bf{0.2 - 10 keV} & $40.4\pm11.4$ & $4.19\pm2.05$ & \\
  0.2 - 0.5 keV & $<22.6$ & $<10.19$ & \\
  0.5 - 2.0 keV & $24.0\pm7.8$ & $1.43\pm0.5$ & \\
  2.0 - 5.0 keV & $<8.9$ & $<26.84$ & \\
  5.0 - 10 keV & $<1.6$ & $<105.23$ & \\
  \hline
  \multicolumn{4}{c}{VIK J0305$-$3150 ($z=6.605$)}\\
  \hline  
  \bf{0.2 - 10 keV} & $<13.3$ & $<5.40$ & \\
  0.2 - 0.5 keV & $<1.9$ & $<2.04$ & \\
  0.5 - 2.0 keV & $<13.5$ & $<2.54$ & \\
  2.0 - 5.0 keV & $<1.9$ & $<5.11$ & \\
  5.0 - 10 keV & $<9.1$ & $<28.57$ & \\
  \hline
  \multicolumn{4}{c}{PSO J338.2298+29.5089 ($z=6.658$)}\\
  \hline  
  \bf{0.2 - 10 keV} & $<22.6$ & $<5.89$ & \\
  0.2 - 0.5 keV & $<9.7$ & $<2.32$ & \\
  0.5 - 2.0 keV & $<19.3$ & $<3.41$ & \\
  2.0 - 5.0 keV & $<12.5$ & $<5.02$ & \\
  5.0 - 10 keV & $<7.1$ & $<17.02$ & \\
  \hline
  \multicolumn{4}{c}{VIK J0109$-$3047 ($z=6.747$)}\\
  \hline  
  \bf{0.2 - 10 keV} & $<10.0$ & $<16.73$ & \\
  0.2 - 0.5 keV & $<2.6$ & $<6.03$ & \\
  0.5 - 2.0 keV & $<10.9$ & $<6.8$ & \\
  2.0 - 5.0 keV & $<5.5$ & $<18.39$ & \\
  5.0 - 10 keV & $<1.3$ & $<60.40$ & \\
\hline\end{tabular}
}
  
  \textbf{Note}: The upper limits quoted in this table correspond to the $3\sigma$ upper limits and are used only if the nets counts or fluxes are smaller than their $3\sigma$ error. But only the quasars with a given value in the "Det" column have an X-ray detection.
\end{table}


\section{Discussion}
\label{sec:disc}





For the full sample of nine quasars 
we derived a set of optical and X-ray properties (see Table \ref{tab:XrayOpt}):
\begin{itemize}
	\item[--] The absolute magnitude at the rest-frame wavelength of $\lambda=1450\mathrm{\AA}$ ($\mathrm{M_{1450}}$) was computed by extrapolating the J-band magnitude to the monochromatic magnitude at $\lambda=1450\mathrm{\AA}$ ($m_{1450}$) assuming a power-law fit of the continuum in the UV-optical $f_\nu \propto \nu^{\alpha_\nu}$ with $\alpha_\nu=-0.3$ \citep{Selsing16}.
	\item[--] The monochromatic flux at the rest-frame wavelength of $2500\mathrm{\AA}$ ($f_{2500\mathrm{\AA}}$) was derived from $m_{1450}$ assuming also a power-law slope of $\alpha_\nu=-0.3$.
        \item[--] The X-ray luminosity in the rest-frame 2-10 keV band was computed from the count rate obtained by the \texttt{eregionanalyse} task on the 2-10 keV image (as done in Section \ref{sec:data} for the other energy ranges) and then K-correction was applied assuming a photon index $\Gamma=1.9$. For the two detected quasars VDES J0224-4711 and PSO J159.2257-02.5438 the flux was obtained directly from their X-ray spectrum. Upper limits are used if the luminosity is smaller than three times its error (which is the case for all the non-detected sources) and are at the $3\sigma$ level.
        \item[--] The hardness ratio is defined as $HR=(H-S)/(H+S)$ where S and H are the number of counts in the soft (0.5-2.0 keV) and hard (2.0-5.0 keV) bands.
        \item[--] The optical-to-X-ray slope ($\alpha_{ox}$) is defined as
           \begin{equation}
               \alpha_{ox} = 0.3838 \times \log \left( \frac{f_{2\mathrm{keV}}}{f_{2500\mathrm{\AA}}} \right)
           \end{equation}
        where $f_{2500\mathrm{\AA}}$ is the monochromatic flux at the rest-frame wavelength of $2500\mathrm{\AA}$ as defined above and the rest-frame monochromatic flux at 2 keV ($f_{2\mathrm{keV}}$) was inferred from the flux in the 0.2-10 keV range assuming a photon index $\Gamma=1.9$ Upper-limits for $\alpha_{ox}$ are used if the flux at 2 keV is smaller than three times its error.
\end{itemize}

\begin{table*}
  \centering
  \caption{Summary of X-ray and optical properties of the DES and new $z>6.4$ X-ray quasars}
  \label{tab:XrayOpt}
  \begin{threeparttable}
     \begin{tabular}{lccccccc}
\hline
  \multicolumn{1}{c}{\multirow{2}{*}{Object}} &
  \multicolumn{1}{c}{\multirow{2}{*}{z}} &
  \multicolumn{1}{c}{\multirow{2}{*}{$J_{AB}$}} &
  \multicolumn{1}{c}{\multirow{2}{*}{$\rm{M_{1450}}$}} &
  \multicolumn{1}{c}{$f_{2500\rm{\AA}}$} &
  \multicolumn{1}{c}{$L_{2-10\rm{keV}}$} &
  \multicolumn{1}{c}{\multirow{2}{*}{HR}} &
  \multicolumn{1}{c}{\multirow{2}{*}{$\alpha_{ox}$\tnote{a}}} \\
   & & & & \multicolumn{1}{c}{($10^{-28}\;\rm{erg\;cm^{-2}\;s^{-1}\;Hz^{-1}}$)} & \multicolumn{1}{c}{($10^{45}\;\rm{erg\;s^{-1}}$)} & & \\
\hline
  VDES J0224$-$4711 & 6.526 & $19.75\pm0.06$ & $-27.03$ & $5.17\pm0.29$ & $2.92\pm0.43$ & $-0.56\pm0.11$ & $-1.95\pm0.03$\\
  VDES J0244$-$5008 & 6.724 & $20.23\pm0.13$ & $-26.61$ & $3.34\pm0.38$ & $<4.37$ & $-$ & $<-1.89$\\
  VDES J0020$-$3653 & 6.834 & $20.4\pm0.10$ & $-26.47$ & $2.88\pm0.26$ & $<4.76$ & $-$ & $-2.07\pm0.07$\\
  \hline\hline
  PSO J159$-$02 & 6.38 & $20.13\pm0.05$ & $-26.62$ & $3.64\pm0.16$ & $2.59\pm0.82$ &  $-0.52\pm0.12$ & $-1.88\pm0.03$ \\
  CFHQS J0210$-$0456 & 6.438 & $22.28\pm0.27$ & $-24.98$ & $0.50\pm0.11$ & $<10.00$ & $-$ & $<-1.49$\\
  PSO J036+03 & 6.541 & $19.46\pm0.10$ & $-27.33$ & $6.78\pm0.60$ & $<17.62$ & $-0.58\pm0.36$ & $-2.13\pm0.06$\\
  VIK J0305$-$3150 & 6.605 & $20.71\pm0.09$ & $-26.10$ & $2.15\pm0.18$ & $<4.28$ & $-$ & $<-1.90$\\
  PSO J338+29 & 6.658 & $20.74\pm0.09$ & $-26.08$ & $2.09\pm0.17$ & $<3.80$ & $-$ & $<-1.88$\\
  VIK J0109$-$3047 & 6.747 & $21.32\pm0.14$ & $-25.53$ & $1.23\pm0.15$ & $<12.59$ & $-$ & $<-1.62$\\
\hline
\end{tabular}

     \begin{tablenotes}
         \small
         \item [a] Reminder: A non-upper limt on the $\alpha_{ox}$ slope does not imply a significant detection. Only VDES J0224-4711 and PSO J159-02 are detected in the X-rays. Upper-limits are used only if the flux at 2keV is smaller that its 3$\sigma$ error.
     \end{tablenotes}
  \end{threeparttable}
\end{table*}
 
\begin{figure}
   \centering
   \includegraphics[width=\columnwidth]{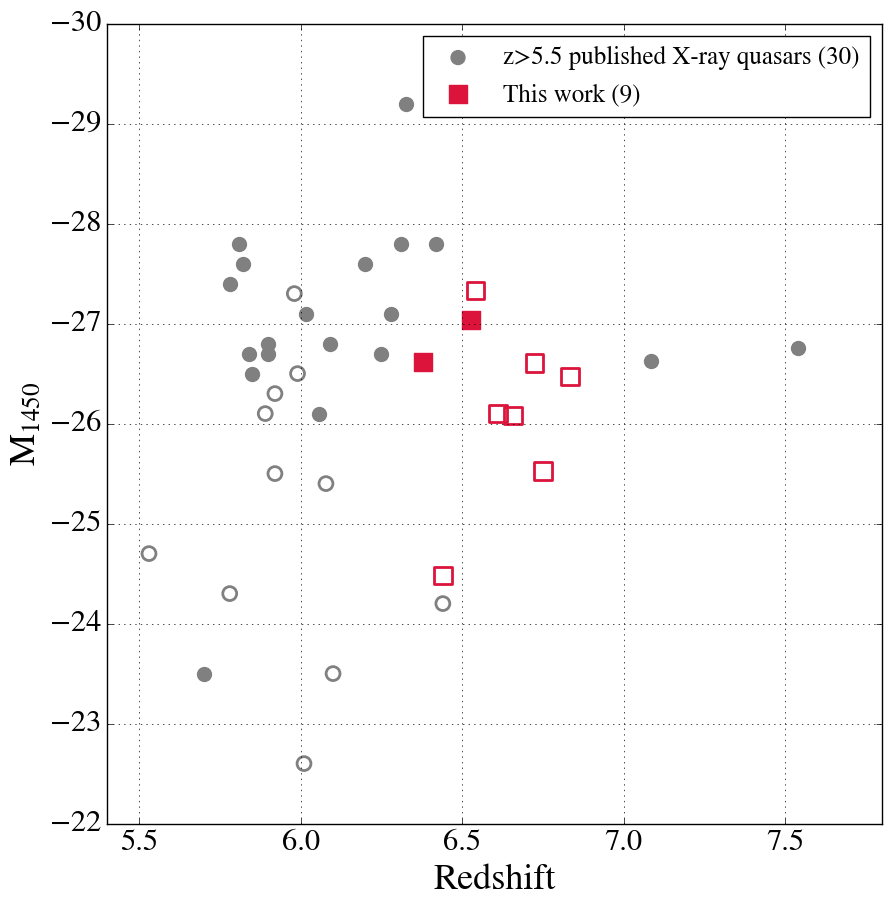}
   \caption{Absolute magnitude $M_{1450}$ versus redshift of all the X-ray observed quasars at $z>5.5$, including our three quasars from DES (red squares), the newly observed Pan-STARRS quasars by N. Schartel (blue diamonds) and the published quasars as compiled by \citet{Nanni17} plus those by \citet{Nanni18} and \citet{Banados18b} (grey circles). The open symbols correspond to undetected X-ray sources and filled symbols are for quasars detected in the X-rays. The new sample from this work allows to fill the gap of X-ray observed quasars between $6.4<z<7.0$}
   \label{fig:results_m1450}
\end{figure}

\begin{figure*}
   \centering
   \begin{tabular}{cc}
       \includegraphics[width=\columnwidth]{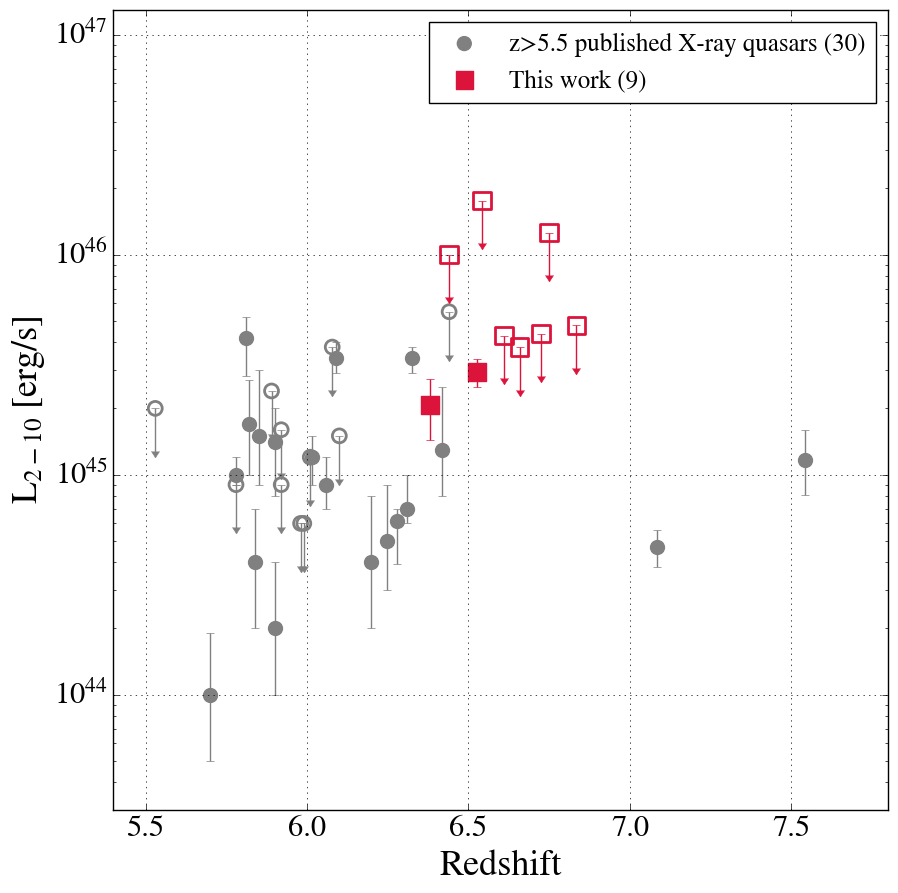}&
       \includegraphics[width=\columnwidth]{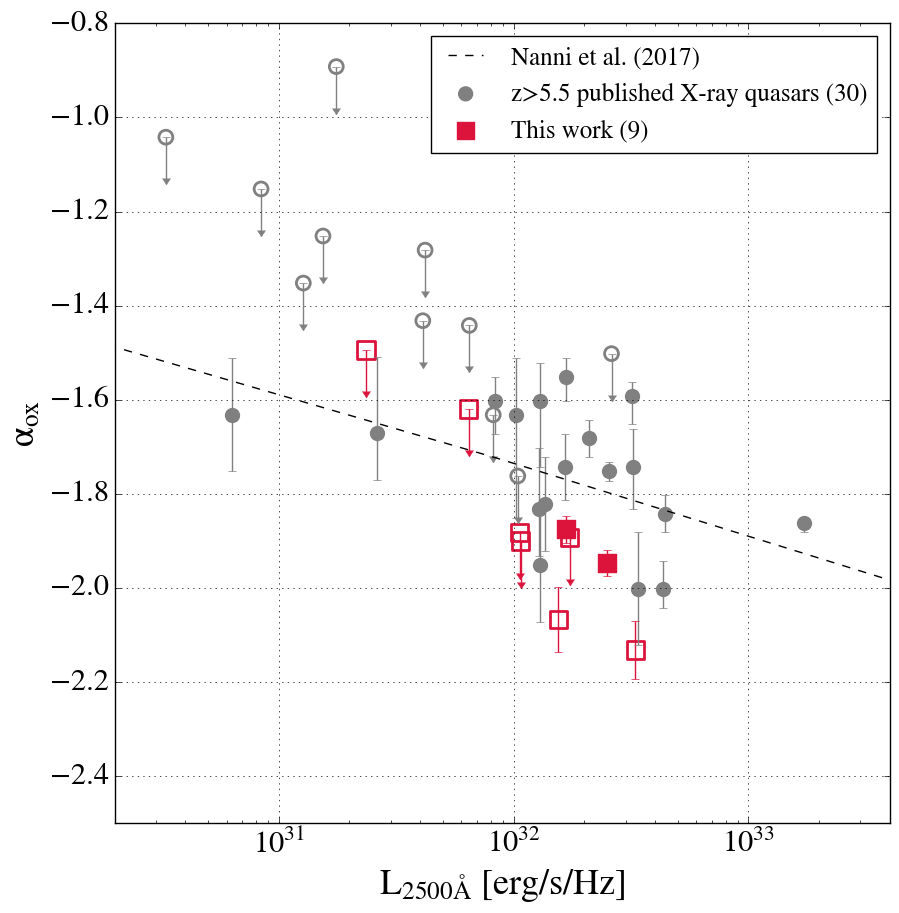}\\
   \end{tabular}
   \caption{X-ray and optical properties of all the X-ray observed quasars at $z>5.5$. For a description of symbols see Figure \ref{fig:results_m1450}. \emph{Left panel}: X-ray luminosity in the rest-frame 2-10 keV band $L_{2-10\mathrm{keV}}$ versus redshift. The two detected new quasars from this work have an X-ray luminosity in the upper range of the distribution of $L_{2-10\mathrm{kev}}$ for $z>5.5$ quasars, with an X-ray luminosity 6 times times larger than ULAS J1120+0641 ($z=7.08$). \emph{Right panel}: Correlation between the optical-to-X-ray slope $\alpha_{ox}$ and the luminosity at the rest-frame $\lambda=2500\mathrm{\AA}$ ($L_{2500\mathrm{\AA}}$). The dashed line corresponds to the best fit relation from \citet{Nanni17}, and the quasars from this work are broadly consistent with this expectation. \textbf{Note}: A non-upper limit on the $\alpha_{ox}$ slope does not imply a detection, only the filled symbols correspond to significant ($4\sigma$) detections.}
   \label{fig:results_plts}
\end{figure*}

We used the compilation of X-ray observations of quasars at $z>5.5$ by \citet{Nanni17} plus the new analysis by \citet{Nanni18} for SDSS J1030+0524 (z=6.28) and the observation of ULAS J1342+0928 (z=7.54) by  \citet{Banados18b} to compare with the properties of our new sample. 
This new sample allows us to fill the gap between $6.4<z<7.0$ of observed high-redshift quasars in the X-rays.

The magnitude $M_{1450}$ versus redshift distribution of the X-ray observed $z>5.5$ quasars is shown in Figure \ref{fig:results_m1450}. The median absolute magnitude at $\lambda=1450\mathrm{\AA}$ is $\rm M_{1450} = -26.1$ for 
the nine newly observed quasars, which is similar to the median value ($\rm M_{1450} = -26.7$) for the sample of previously known $z>5.5$ quasars including the two quasars at $z>7.0$.
Only one quasar with $M_{1450}>-25$ has been previously detected in the X-rays 
\citep[RD J1148+5253, $z=5.7$;][]{Gallerani17}. This quasar was a serendipitous
detection in the 78ks \emph{Chandra} observation of the $z=6.4$ quasar SDSSJ1148+5251. 
It has about 4 times more exposure time than the other X-ray detected high-redshift quasars, 
and only $\sim3$ counts were detected. For comparison, the lowest UV luminosity source in 
our sample, CFHQS J0210$-$0456 ($\mathrm{M_{1450}=−24.98}$), has an upper limit of 14 counts in the 0.2-10 keV band.

For the two X-ray detected quasars, we found that their X-ray spectra are well fitted with a standard AGN model which consists of a power-law modified by Galactic absorption. The power-law has a photon index $\Gamma \sim 1.9$ which is consistent with value observed at $z\sim 6$ quasars by \citet{Nanni17} and also with lower redshift quasars, as firstly observed by \citet{Nandra97}. Another indicator of the X-ra spectral shape is the hardness ratio. In the absence of obscuration low and high redshift quasars should appear soft in the X-rays and are expected to have a low hardness ration with $\mathrm{HR}\sim -0.5$ \citep{Wang04}. For the sources with enough counts we obtained an hardness ratio $HR\sim-0.55$ (see Table \ref{tab:XrayOpt}), similar to what is expected for high-redshift quasars and also what have been observed in $z\sim6$ quasars \citet{Nanni17}.

On the left panel of Figure \ref{fig:results_plts} we show the X-ray luminosity in the rest-frame 2-10 keV band ($L_{2-10\mathrm{kev}}$). The new detected sources in our work are amongst the most luminous X-ray quasars at $z>5.5$ with $L_{2-10\mathrm{keV}} \sim 2-4\times10^{45}\;\mathrm{erg\;s^{-1}}$. VDES J0224$-$4711 which is the only DES quasar detected in the X-rays has a rest-frame $2-10$ keV luminosity of $L_{2-10\mathrm{keV}} = (2.92\pm0.43)\times10^{45}\;\mathrm{erg\;s^{-1}}$, 6 times and 2.5 times larger than the ULAS J1120+0641 ($z=7.08$) and ULAS J1342+0928 ($z=7.54$) respectively. The only other quasar detected is PSO J159$-$02 with $L_{2-10\mathrm{keV}} = (2.08\pm0.65)\times10^{45}\;\mathrm{erg\;s^{-1}}$, i.e. a value similar to VDES J0224$-$4711. The other quasars have upper limits in their X-ray luminosity with $L_{2-10\mathrm{keV}} < 2\times10^{46}\;\mathrm{erg\;s^{-1}}$.

The optical-to-X-ray slope $\alpha_{ox}$ computed for the quasars in our sample is consistent with the 
observed $\alpha_{ox}-L_{2500\mathrm{\AA}}$ correlation inferred by \citet{Nanni17} who added 29 quasars with $z>5.5$ to the a sample of a thousand lower redshift quasars (see the right panel of Figure \ref{fig:results_plts}). These luminous quasars at $z>6.4$ follow the trend of decreasing $\alpha_{ox}$ as $L_{2500\mathrm{\AA}}$ increases, as found in previous studies \citep{Steffen06, Lusso10} for quasars at lower redshift. To be consistent we re-computed the $\alpha_{ox}$ and $L_{2500\mathrm{\AA}}$ from \citet{Nanni17} assuming a power-law slope in the UV-optical of $\alpha_\nu=-0.3$ (while $\alpha_\nu=-0.5$ is used by \citet{Nanni17}). Changing the slope only has a small effect, as $L_{2500\mathrm{\AA}}$ increases by $\sim10\%$ and $\alpha_{ox}$ is larger by 0.018.

\begin{figure}
   \centering
   \includegraphics[width=\columnwidth]{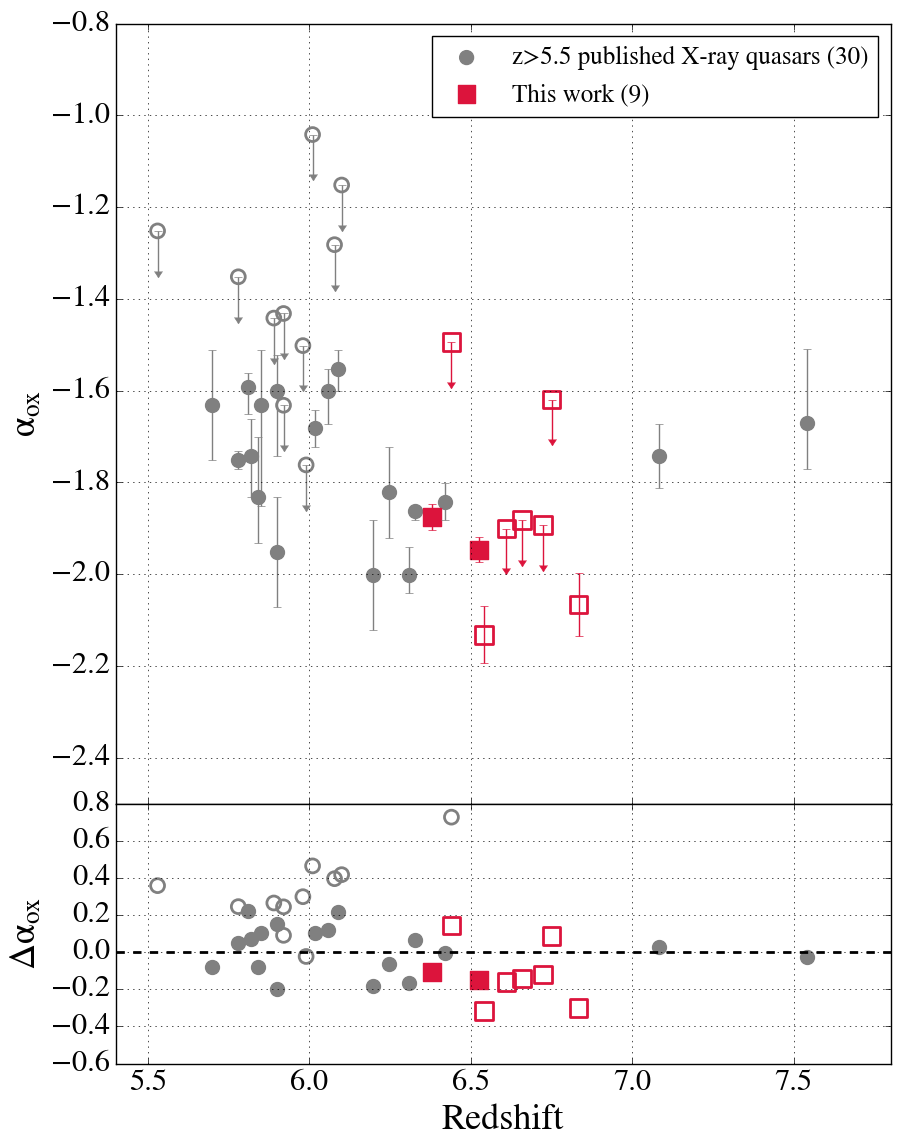}
   \caption{\emph{Top panel}: Optical-to-X-ray slope $\alpha_{ox}$ versus redshift. For a desciption of the symbols see Figure \ref{fig:results_m1450}. \emph{Bottom panel}: Difference between the computed $\alpha_{ox}$ and the predicted $\alpha_{ox}$ from $L_{2500\mathrm{\AA}}$ using the correlation by \citet{Nanni17} ($\Delta\alpha_{ox}$). We do not observed obvious correlation between $alpha_{ox}$ and the redshift as state by previous works.}
   \label{fig:alphaox_z}
\end{figure}

Previous studies of $\alpha_{ox}$ in quasars \citep{Strateva05, Shemmer05, Just07, Lusso10} have tested for  possible redshift dependence of $\alpha_{ox}$ but they did not find any significant correlation between the two. 
The apparent correlation between $\alpha_{ox}$ and the redshift (i.e. $\alpha_{ox}$ seems to decrease as the redshift increases from $z\sim0$ up to $z\sim6$) for quasars with broad range of luminosity $10^{27}<L_{2500\mathrm{\AA}}<10^{33}\;\mathrm{erg\;s^{-1}\;Hz}$ is in fact only an artifact due 
to the effect of the optical luminosity.
The X-ray quasars at $z>5.5$ span a narrower range of optical luminosity ($10^{30}<L_{2500\mathrm{\AA}}<10^{33}\;\mathrm{erg\;s^{-1}\;Hz}$) and we do not observe any obvious evolution 
of $\alpha_{ox}$ with redshift (see Figure \ref{fig:alphaox_z}; top panel), especially when correcting for the effect of the luminosity (see Figure \ref{fig:alphaox_z}; bottom panel).

\section{Conclusion}
 
 We present \emph{XMM-Newton} X-ray observations and analysis of three $z>6.5$ quasars 
previously discovered in the Dark Energy Survey and the X-ray analysis of six other  quasars with
$6.438 < z < 6.747$ from the \emph{XMM-Newton} public archive.
 The analysis of these sources show that two quasars VDES J0224$-$4711 ($z=6.526$) and PSO J159$-$02 ($z=6.38$)
 are detected in the X-rays 0.2-10 keV energy band. VDES J0224$-$4711 is detected with a significance of $9\sigma$ and $\sim174$ net counts corresponding to a flux of $f_{2-10\mathrm{keV}} \sim 12\times10^{-15}\;\mathrm{erg\;cm^{-2}\;s^{-1}}$. PSO J159$-$02 is detected with a significance of $8\sigma$, and $\sim127$ net counts, corresponding to a flux of $f_{2-10\mathrm{keV}} \sim 11\times10^{-15}\;\mathrm{erg\;cm^{-2}\;s^{-1}}$. 
 
 VDES J0224$-$4711 is also detected individually in the 0.5$-$2.0 keV soft band and PSO J159$-$02 has individual detection in both the 0.5$-$2.0 keV and 2.0$-$5.0 keV bands. The X-ray spectra are well fitted with a power-law with a photon index $\Gamma\sim1.82^{+0.29}_{-0.27}$ and $\Gamma\sim1.94^{+0.31}_{-0.29}$ respectively, similarly to what is observed in other quasars at lower redshifts \citep{Nanni17}. 
 For the seven other quasars undetected in the X-ray (i.e. detection significance $<4\sigma$) we were still able to estimate X-ray fluxes (or $3\sigma$ upper limits) by extracting the number of counts in a circular region around the source position.
 
 The X-ray observations of these 9 quasars allow us to fill the gap in the 
 redshift distribution of 
 high redshift quasars observed in the X-rays for $6.5<z<7.0$. Prior to this work, only two quasars ULAS J1120+0641 and ULAS J1342+0928 had X-ray observations among the known quasars at $z>6.5$. We find that the quasars in our sample have similar $M_{1450}$ magnitudes compared to X-ray observed quasars at $z>5.5$ with the exception of CFHQS J0210$-$0456 which has one the faintest magnitude ($M_{1450}\sim-24.5$). Furthermore, for the quasars with discrete measurements of $L_{2-10\mathrm{kev}}$, we find that they are among the brightest X-ray quasars at $z>5.5$ with $L_{2-10\mathrm{keV}} \sim 2-4\times10^{45}\;\mathrm{erg\;s^{-1}}$, with VDES J0224$-$4711 and PSO J159$-$02 - the two quasars with clear X-ray detections - being 6 times an 2.5 times larger than the ULAS J1120+0641 ($z=7.08$) and ULAS J1342+0928 ($z=7.54$). Finally the optical-to-X-ray slope $\alpha_{ox}$ we get for the quasars in our sample is consistent with the $\alpha_{ox}-L_{2500\mathrm{\AA}}$ anti-correlation inferred by \citet{Nanni17} based on low and high-redshift quasars.
 




\section*{Acknowledgements}

EP, RGM acknowledge the support of UK Science and Technology research Council (STFC). RGM also acknowledges support by ERC Advanced Grant 320596 \emph{The Emergence of Structure during the Epoch of reionization}. This work was made possible thanks to observations obtained with XMM-Newton, an ESA science mission with instruments and contributions directly funded by ESA Member States and NASA.



\bibliographystyle{mnras}
\bibliography{refs} 


\bsp	
\label{lastpage}
\end{document}